\documentclass{emulateapj}
\shorttitle{ULUV source in NGC 6946}
\shortauthors{Kaaret et al.}

\begin{document}

\title{Direct Detection of an Ultraluminous Ultraviolet Source}

\author{Philip Kaaret\altaffilmark{1}, Hua Feng\altaffilmark{2,1}, Diane S. Wong\altaffilmark{3,1}, and Lian Tao\altaffilmark{2}}

\altaffiltext{1}{Department of Physics and Astronomy, Van Allen Hall, University of Iowa, Iowa City, IA 52242 USA}

\altaffiltext{2}{Department of Engineering Physics and Center for Astrophysics, Tsinghua University, Beijing 100084, China}

\altaffiltext{3}{Astrophysics Department, 601 Campbell Hall, University of California, Berkeley, CA 94720 USA}

\begin{abstract}

We present Hubble Space Telescope observations in the far UV of the ultraluminous X-ray source in NGC 6946 associated with the optical nebula MF 16.  Both a point-like source coincident with the X-ray source and the surrounding nebula are detected in the FUV.  The point source has a flux of $5 \times 10^{-16} \rm \, erg \, s^{-1} \, cm^{-2} \, \AA^{-1}$ and the nebula has a flux of $1.6 \times 10^{-15} \rm \, erg \, s^{-1} \, cm^{-2} \, \AA^{-1}$, quoted at 1533 \AA\  and assuming an extinction of $A_V = 1.54$.  Thus, MF 16 appears to host the first directly detected ultraluminous UV source (ULUV). The flux of the point-like source is consistent with a blackbody with $T \approx 30,000$~K, possibly from a massive companion star, but this spectrum does not create sufficient ionizing radiation to produce the nebular He{\sc ii} flux and a second, hotter emission component would be required.  A multicolor disk blackbody spectrum truncated with an outer disk temperature of $\sim 16,000$~K provides an adequate fit to the FUV, B, V, I, and He{\sc ii} fluxes and can produce the needed ionizing radiation.  Additional observations are required to determine the physical nature of the source.

\end{abstract}


\section{Introduction}

Ultraluminous X-ray sources (ULXs) are bright, irregularly variable, non-nuclear, X-ray sources with apparent luminosities exceeding the Eddington limit for a $20 M_{\odot}$ compact object.  ULXs could represent a new class of black holes with masses intermediate between stellar-mass and supermassive black holes \citep{Colbert99,Makishima00,Kaaret01} or may be stellar-mass black holes with unusually high accretion rates and/or beamed emission.  Over the past few years, optical and radio nebulae have been found which are spatially coincident and likely physically related to ULXs \citep{Pakull02,Roberts03,Kaaret04,Lang07}.  Study of nebulae associated with ULXs may enable us to determine the total radiation and particle fluxes from the ULXs if the nebulae are continuously powered by photoionization or particle flows.

The nebula MF 16 in the nearby (5.1 Mpc) and nearly face-on spiral galaxy NGC 6946 has been described as an `ultraluminous supernova remnant complex' \citep{Blair01}.  The object is unusually luminous for a SNR in optical line emission, e.g.\ the H$\alpha$ luminosity is $2 \times 10^{38} \rm \, erg \, s^{-1}$, and in X-rays \citep{Schlegel94}. \citet{Roberts03} found that the X-ray source is variable and, thus, not diffuse X-ray emission associated with a SNR, but rather emission from an X-ray binary containing a compact object.  MF 16 which had been previously thought of as a SNR is, instead, a nebula surrounding an X-ray binary.

MF 16 shows many lines consistent with shock emission.  However, several of the lines (e.g.\ [O {\sc iii}], H$\alpha$, and [N {\sc ii}]) contain a narrow component, generally associated with photoionization instead of shocks.  \citet{Blair01} concluded that unusually strong photoionization is needed, in addition to shocks, to explain the spectrum, but did not determine the source of ionizing photons.  They also reported detection of He{\sc ii} line emission, but did not discuss the origin of the line.

\citet{Abolmasov08} studied the optical emission from MF 16 with particular attention to the He{\sc ii} $\lambda 4686$ line.  Their modeling shows that most of the power in the optical line emission comes from photoionization, although a component from shocks is still important.  They find that an ultraviolet source more luminous than the observed ULX is needed to power the emission and that the UV source has a blackbody temperature of $10^{5.15 \pm 0.10}$~K and a luminosity  $1.2 \times 10^{40} \rm \, erg \, s^{-1}$.  The luminosity is a lower bound on the source luminosity if the photoionized nebula is density bounded.

The UV source proposed by \citet{Abolmasov08} should be bright in the FUV.  We obtained FUV observations using the solar blind channel (SBC) of the Advanced Camera for Surveys (ACS) on the Hubble Space Telescope (HST) to search for the predicted FUV emission.  We describe the observations and analysis in \S 2.  We discuss the implications in \S 3.

\begin{figure*} \centerline{\includegraphics[width=4.75in]{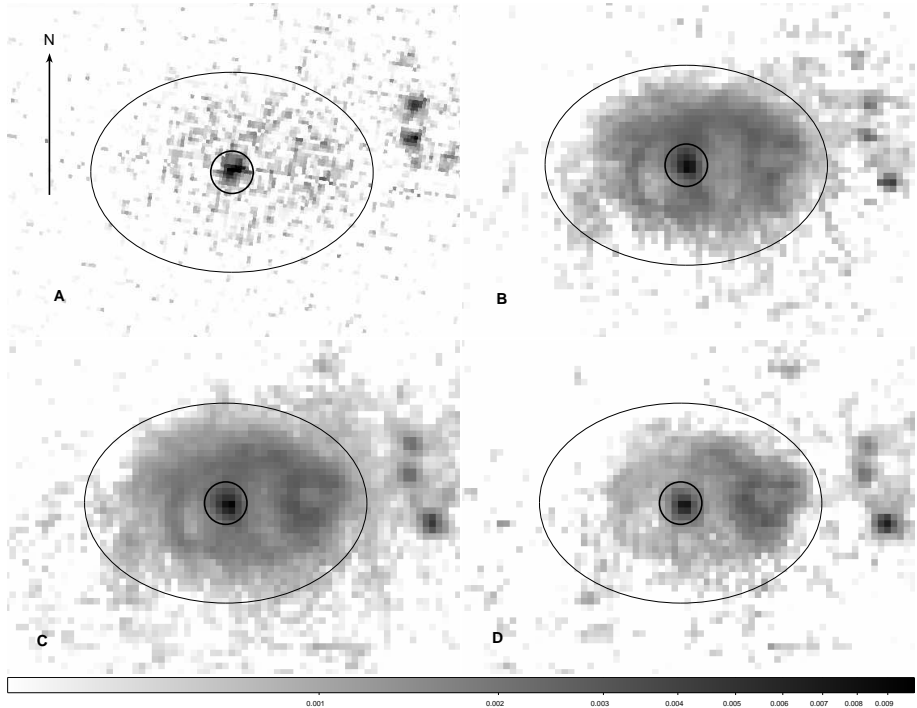}}
\caption{Images of MF 16.  The panels show A: FUV (F140LP filter), B: B-band (F450W filter), C: V-band (F555W filter), and D: I-band (F814W filter).  The small circles mark the counterparts to the X-ray source and have radii of $0.15\arcsec$.  The ellipses mark the extent of the nebula with a major diameter of $2.0\arcsec$ and a minor diameter of $1.4\arcsec$.  The arrow in panel A points North and is $1\arcsec$ long. \label{images}} \end{figure*}

\section{Observations}\label{obs}

Observations were performed on 2008 May 01 using the ACS/SBC on HST centered on the position of the X-ray source in MF 16 (GO program 11200, PI: Kaaret).  The spectral element used was the F140LP long-pass filter which covers the 1350--2000\AA\ band.  This filter was chosen because it has a much smaller background sky rate, blocking out geocoronal Lyman $\alpha$ emission, than the filters that extend to shorter wavelengths.  The exposure time was 2760 seconds.

We also extracted several images of the field from the HST archives, these include an ACS Wide Field Camera (WFC) image in the F814W (I-band) filters obtained on 2004 July 29 and the Wide Field and Planetary Camera 2 (WFPC2) images in the F450W (B-band), F555W (V-band), and F814W (I-band) filters all obtained on 2001 June 8.  For the ACS data, we used the dithered images directly from the archive.  For the WFPC2 data, we combined the 4 images obtained in each filter and performed cosmic ray rejection using {\tt multidrizzle} in Pyraf.

We re-analyzed an archival observation made with the Advanced CCD Imaging Spectrometer (ACIS) on the Chandra  X-ray observatory on 2001 September 07 UT (ObsID 1043).  This observation was chosen because it had the longest exposure time, 59~ks, of those available in the archive, allowing the deepest search for X-ray counterparts to optical sources as needed for the X-ray/optical image alignment.  The {\it Chandra} data were subjected to standard calibration using CIAO v4.0: applying the aspect solution, generating an instrument map weighted by an absorbed power law, and generating an exposure map.  There were no background flares, so the entire exposure time was used. 

\subsection{Astrometry}

The Two-Micron All Sky Survey (2MASS,\citealt{Skrutskie06}) catalog was chosen for absolute astrometry correction over other catalogs, e.g., USNO-B1.0, because of its superior accuracy.  Because the SBC has a small (34\farcs6 x 30\farcs8) field of view, there are no 2MASS stars on our image.  Instead, absolute astrometry was performed  using the WFC/F814W image, and then the SBC and Chandra images were aligned to the WFC/F814W image.  

Using the Graphical Astronomy and Image Analysis (GAIA) tool, we detected stars on the WFC/F814W image and then searched for point-like counterparts with magnitudes $K < 15$ in the 2MASS catalog, finding 20 matches.  These matches were used to compute a new aspect solution for the F814W.  The fit rms was 1.9 pixels or $0.9\arcsec$.  Eighty-nine point sources were found in the SBC image, of which 11 are also appear in the F814W image.  These 11 matches were used to transfer the aspect solution to the SBC image.

We found 40 X-ray sources within the field of the F814W image by running the {\tt wavdetect} tool in CIAO on a 0.3--8 keV band image.  Of these 40, 3 have counterparts in the WFC image (excluding the X-ray source in MF 16).  However, one of the objects is extended and another is faint and has a large point-spread function due to its off-axis position in the Chandra image.  So the one bright, point source was used to correct the astrometry.  First, the positions found from {\tt wavdetect} were corrected by subtracting the systematic error found by running a Marx simulation of the Chandra telescope.  The Chandra image was then translated, by $0.2\arcsec$ in right ascension and $-0.3\arcsec$ in declination, to match the F814W image.  The corrected position of the X-ray source in MF 16 is R.A.=20:35:00.752, decl.=+60:11:30.55 (J2000).  From the {\tt wavdetect} position errors, we estimate the 90\% error radius to be $0.4\arcsec$.

\begin{deluxetable}{lcc}
\tablewidth{0pt}
\tablecaption{Photometry\label{photometry}}
\tablehead{
Filter & Count rate & Flux for $A_V = 1.54$ \\
       & (c/s)      & ($\rm erg \, cm^{-2} \, s^{-1} \, \AA^{-1}$)}
\startdata
F140LP (FUV) & $0.24 \pm 0.04$ & $(4.8 \pm 0.7) \times 10^{-16}$ \\
F450W  (B)   & $0.41 \pm 0.06$ & $(2.8 \pm 0.4) \times 10^{-17}$ \\
F555W  (V)   & $0.69 \pm 0.10$ & $(1.3 \pm 0.2) \times 10^{-17}$ \\
F814W  (I)   & $0.33 \pm 0.05$ & $(2.7 \pm 0.4) \times 10^{-18}$ 
\enddata
%\tablecomments{Hello}
\label{rates}
\end{deluxetable}

\subsection{Photometry}

Figure~\ref{images} shows images of MF 16 in the FUV (F140LP filter), B-band (F450W filter), V-band (F555W filter), and I-band (F814W filter).  The nebula and a counterpart to the X-ray source are clearly visible in each band.  The nebula appears extended along the East-West direction.  The ellipse drawn on the figure has a major diameter of $2.0\arcsec$ and a minor diameter of $1.4\arcsec$ corresponding to a physical size of 50~pc by 35~pc at a distance of 5.1~Mpc \citep{deVaucouleurs79}.

We performed aperture photometry on the X-ray source counterparts, extracting counts from a circle with a radius of $0.15\arcsec$.  We use the WFPC2 F814W image, rather than the ACS/WFC one, since the source may be variable and the three WFPC2 images were obtained over a span of less than 5 hours.  We subtracted the nebular flux, scaled by area, from the point source flux.  If, instead, we use a background region external to the nebula, i.e.\ retain the nebular contribution within the $0.15\arcsec$ extraction circle, the FUV counts increase by 11\% and the BVI counts increase by $\sim 30\%$.  We relied on the {\tt synphot} package in IRAF for the aperture correction for the ACS data.  For the WFPC2, we determined the aperature correction (from a radius of $0.15\arcsec$ to $1.0\arcsec$) using the star USNO B1.0 1501-0283184 also located on the planetary camera chip.  The same star was also used to check our photometry in the B and I bands.  Nebular-background subtracted count rates for the point source for each continuum filter are given in Table~\ref{rates}.  We estimate a 15\% error on the source count rates, primarily due to uncertainties in the background subtraction (due to the presence of the nebula).  The background-subtracted count rate in the F140LP filter for the nebula, excluding the point source, is 0.78~c/s.  For the point source, we find equivalent magnitudes of $B=22.73$ and $V=22.70$.  The V magnitude is in good agreement with $V=22.64$ as quoted by \citet{Blair01}, while our B-magnitude is somewhat brighter than their value of $B=23.10$.

We modeled the intrinsic source flux as a power-law, $F_{\lambda} \propto \lambda^{-2}$ and used the {\tt calcphot} command in the synphot package to relate the measured count rates to intrinsic source fluxes.  We note that the inferred fluxes change by less than 1\% if we change the power-law spectral index to $F_{\lambda} \propto \lambda^{-4}$.  The filter pivot wavelengths are 1533 \AA\ for F140LP,  4557 \AA\ for F450W, 5443 \AA\ for F555W, and 7996 \AA\ for F814W.  The Galactic interstellar extinction along the line of sight to MF16 is $A_V = 1.14$~mag \citep{Schlegel98}.  Reported values of the additional, intrinsic extinction, mostly found by measuring decrements between Balmer lines in the nebular spectrum, vary from 0.2 to 0.66 magnitudes \citep{Blair01,Roberts03,Abolmasov08}.  Following \citet{Abolmasov08}, we adopt an extinction of $A_V = 1.54$~mag and the extinction curve of \citet{Cardelli89}.  The point source fluxes are given in Table~\ref{rates} for $A_V = 1.54$~mag.  The lower extinction value discussed by \citet{Abolmasov08}, $A_V = 1.34$~mag, decreases the intrinsic flux by 41\% in the FUV, 22\% in B, 18\% in V, and 11\% in I.  The lower extinction may be valid for the point source, if source preferentially ionizes the nearby regions of the nebula.  Adopting an extinction at the high end of the reported values, $A_V = 1.8$~mag \citep{Blair01,Roberts03}, increases the intrinsic flux by 84\% in the FUV, 33\% in B, 26\% in V, and 15\% in I.

\begin{figure}[t] \centerline{\includegraphics[width=3.1in]{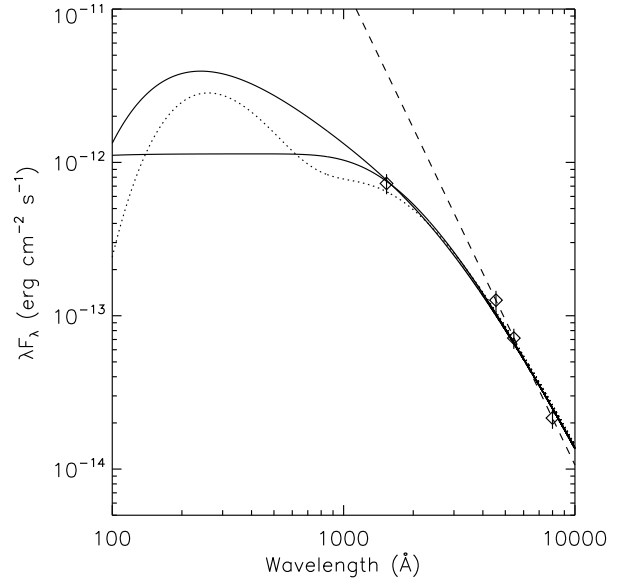}}
\caption{Spectral energy distribution of the point-like source at the center of MF 16.  The dereddened fluxes from Table~\ref{rates} are plotted as diamonds.  The dashed line is a power-law with $F_{\lambda} \propto \lambda^{-4.1}$.  The dotted curve is the sum of two blackbody spectra with temperatures of 140,000~K and 26,000~K.  The solid curves show multicolor disk blackbody spectra.  The curve with higher flux at 200 \AA\ is for $p = 0.75$ and the lower one is for $p=0.5$.  \label{mspec}} \end{figure}

\section{Discussion}

FUV imaging of MF 16 reveals both a point-like source coincident with the X-ray source and a surrounding nebula with an angular extent of $2.0\arcsec$ by $1.4\arcsec$.  The point source has a flux of $5 \times 10^{-16} \rm \, erg \, s^{-1} \, cm^{-2} \, \AA^{-1}$ and the nebula has a flux of $1.6 \times 10^{-15} \rm \, erg \, s^{-1} \, cm^{-2} \, \AA^{-1}$, both quoted at 1533 \AA, corresponding to the pivot wavelength of the F140LP filter and assuming an extinction of $A_V = 1.54$.   The monochromatic luminosity estimate is $\lambda L_{\lambda} = 1.0 \times 10^{40} \rm \, erg \, s^{-1}$, suggesting the presence of an ultraluminous UV source (ULUV).  The measured flux is close to that predicted by \citet{Abolmasov08} near 1000 \AA.

Figure~\ref{mspec} shows the spectral energy distribution of the point-like source at the center of MF 16 using the data from the FUV (F140LP), B (F450W), V (F555W), and I (F814W) bands.  The dereddened fluxes are from Table~\ref{rates}.   We note that the BVI data are quasi-simultaneous, obtained within 5 hours, while the FUV data were obtained 7 years later.  Thus, the non-simultaneity of the observations should be of concern in fitting the data.  However, \citet{Fridriksson08} report that the rms variation in X-ray flux of this source in Chandra observations spread over several years is only 8\%.  Also, the I-band flux from the ACS/WFC measured in 2004 agrees with that measured with the WFPC2 in 2001 within 25\%.  Thus, fitting of these non-simultaneous data appears justified.

A fit of a power-law, $F_{\lambda} = A (\lambda/6000 \rm \AA)^{\alpha}$ to the BVI data dereddened assuming $A_V = 1.54$ gives best fit parameters $\alpha = -4.1 \pm 0.4$ and $A = (8.8 \pm 0.8) \times 10^{-18} \rm \, erg \, s^{-1} \, cm^{-2} \, \AA^{-1}$, see Fig.~\ref{mspec}.  Using $A_V = 1.8$ makes the curve somewhat steeper, $\alpha = -4.4 \pm 0.4$, while $A_V = 1.34$ leads to a slightly shallower slope, $\alpha = -3.9 \pm 0.4$.  In all cases, a good fit is obtained with $\chi^2 < 1$.  The exponent of the the power-law is consistent with that expected for the Rayleigh-Jeans tail of a blackbody spectrum, $\alpha = -4$.

We fit all four data points with a single blackbody spectrum.  This produces a good fit with a temperature $T =$~31000~K and a bolometric luminosity of $3.4\times 10^{39} \rm \, erg \, s^{-1}$.    However, due to the relatively low temperature, the spectrum produces an insignificant flux of radiation capable of ionizing He.  A second source producing radiation at shorter wavelengths is required to produce the He{\sc ii} $\lambda 4686$ luminosity seen from the nebula.  Thus, we added a hotter blackbody with $T =$140,000~K and $L = 1.2 \times 10^{40} \rm \, erg \, s^{-1}$ as suggested by the emission line modeling \citep{Abolmasov08}.  The fit is acceptable, $\chi^2/{\rm DoF} = 3.8/2$, and the cooler blackbody temperature is reduced to $T =$26000~K and its luminosity to $2.2\times 10^{39} \rm \, erg \, s^{-1}$.   These values are for $A_V = 1.54$.  Varying the extinction in the range $A_V = 1.34-1.8$ and the hot blackbody parameters within the certainty range given by \citet{Abolmasov08}, allows temperatures in the range 23,000--32,000~K and luminosities in the range $1.1-4.6 \times 10^{39} \rm \, erg \, s^{-1}$.  These properties are consistent with a high mass star, specifically a late O or early B (O7 to B3) supergiant, that could be the companion star in the binary system.  It would be of interest to obtain an optical and/or UV spectrum of the point source to search for spectral lines that would permit classification of the star, or rule out interpretation in terms of a stellar companion.

The spectra of accretion disks around black holes are often described by a multicolor disk blackbody (MCD) spectrum where the local effective temperature is a decreasing function of radius, usually $T(r) \propto r^{-p}$, between an inner disk radius, $R_{\rm in}$, where the temperature is $T_{\rm in}$ and an outer radius, $R_{\rm out}$, with temperature $T_{\rm out}$.  At long wavelengths, $ hc/\lambda \ll kT_{\rm out}$, the spectrum has the form of a Rayleigh-Jeans tail and at intermediate wavelengths, $kT_{\rm out} \ll hc/\lambda \ll kT_{\rm in}$, the spectrum flattens.  The observed FUV/optical spectrum of the point source is  compatible with this form if the outer disk temperature lies between the FUV and B bands.  The FUV/optical data can constrain MCD model spectra only in regards to the outer disk temperature and the flux in the FUV/optical range.  A constraint on the short wavelength end of the spectrum can be obtained from the ionizing luminosity needed to produce the nebular He{\sc ii} $\lambda 4686$ emission.  The solid lines in Fig.~\ref{mspec} show MCD spectra consistent with the FUV/optical data and with a UV luminosity of $L(20 < \lambda < \rm 228 \AA) \approx 4 \times 10^{39} \rm \, erg \, s^{-1}$ as needed to produce the observed He{\sc ii} line \citep{Abolmasov08}.  We consider two disk temperature profiles: a standard profile with $p = 0.75$ and a profile with $p = 0.5$ as found in spectral fits to a number of ULXs \citep{Vierdayanti06}.  Both fits are acceptable with $\chi^2 < 3.4$.  For $p = 0.75$, the bolometric luminosity is $L_{\rm Bol} = 1.2 \times 10^{40} \rm \, erg \, s^{-1}$ and $T_{\rm out} = 16000 \pm 2000 $~K.  However, the disk is unusually narrow, $R_{\rm out}/R_{\rm in}\sim 40$.  For $p = 0.5$, we find that $L_{\rm Bol} = 8 \times 10^{40} \rm \, erg \, s^{-1}$, $T_{\rm out} = 21000 \pm 2000 $~K, and $R_{\rm out}/R_{\rm in}\sim 6000$.  The true situation may be more complex with $p \sim 0.5$ only in the inner disk and $R_{\rm out}/R_{\rm in}$ intermediate between these two extremes.  It is also be possible to fit the data using models where the outer disk is irradiated.  However, the complexity of irradiation, and the number of parameters in such models, preclude obtaining useful constraints with the available data.  In all cases, the outer disk radius is of order $10^{12} \rm \, cm$.

In summary, two possible scenarios to explain the observational data are that the emission is the sum of a hot blackbody from the compact object and stellar emission from the companion star or that the emission is due to a multicolor disk with an outer disk radius $\sim 10^{12} \rm \, cm$.  The small number of broad band measurements currently available is inadequate to distinguish between these two cases.  Additional observations are needed to determine the physical nature of the source.

\acknowledgments

We acknowledge support from STScI grant HST-GO-11200.  HF acknowledges support from grants No.\ 10903004 and 10978001 from the National Natural Science Foundation of China and grant 2009CB824800 from the 973 Program of China. DSW acknowledges support from the Japan Society for the Promotion of Science. This publication makes use of data products from the Two Micron All Sky Survey, which is a joint project of the University of Massachusetts and the Infrared Processing and Analysis Center/California Institute of Technology, funded by the National Aeronautics and Space Administration and the National Science Foundation.

Facilities: \facility{HST(ACS, WFPC2)}, \facility{Chandra(ACIS)}

\end{document}